\documentclass[conference,10pt]{IEEEtran}
\usepackage{epsfig}
\usepackage{amsmath}
\usepackage{float}
\restylefloat{table}
\usepackage[ruled,vlined]{algorithm2e}
\usepackage{amssymb}
\usepackage{bbm}
\usepackage{setspace}
\usepackage[T1]{fontenc}
\usepackage{textcomp}
\begin{document}

\title{One-bit mmWave MIMO Channel Estimation using Deep Generative Networks}
\author{
\IEEEauthorblockN{
    Akash Doshi and Jeffrey G. Andrews \\}
\IEEEauthorblockA{
		Department of Electrical and Computer Engineering \\
    The University of Texas at Austin, TX 78712, USA\\
    Email: akashsdoshi@utexas.edu, jandrews@ece.utexas.edu}
\thanks{Date of current version: Aug 18, 2022. This work was supported by NVIDIA, Qualcomm Innovation Fellowship (QIF) and the NSF under Grants CNS-2148141 and CCF-2008710. Code and data will be made publicly available at https://github.com/akashsdoshi96/obq-gan-mimo-ce.}
}
\maketitle 
\normalsize

\begin{abstract}
As future wireless systems trend towards higher carrier frequencies and large antenna arrays, receivers with one-bit analog-to-digital converters (ADCs) are being explored owing to their reduced power consumption. However, the combination of large antenna arrays and one-bit ADCs makes channel estimation challenging. In this paper, we formulate channel estimation from a limited number of one-bit quantized pilot measurements as an inverse problem and reconstruct the channel by optimizing the input vector of a pre-trained deep generative model with the objective of maximizing a novel correlation-based loss function. We observe that deep generative priors adapted to the underlying channel model significantly outperform Bernoulli-Gaussian Approximate Message Passing (BG-GAMP), while a single generative model that uses a conditional input to distinguish between Line-of-Sight (LOS) and Non-Line-of-Sight (NLOS) channel realizations outperforms BG-GAMP on LOS channels and achieves comparable performance on NLOS channels in terms of the normalized channel reconstruction error.
\end{abstract}


\section{Introduction}
Channel estimation (CE) in 6G and beyond will be performed at increasingly higher carrier frequencies, leading to an increase in the dimensionality and complexity of the problem due to the associated increase in antenna array sizes at the base station (BS) and user (UE) \cite{rappaport2019wireless}. Conventional sub-6 GHz CE techniques such as least squares (LS) and minimum mean squared error (MMSE) estimators require full rank pilot measurements to recover the channel, hence will not scale to this ``high-dimensional" regime due to the large pilot overhead. In order to reduce the pilot overhead, several compressed sensing (CS) based CE techniques \cite{alkhateeb2014channel,mendez2016hybrid,sun2018joint} have been proposed, in the absence of recieve signal quantization. These same techniques, however, need a significantly larger number of pilots when combined with few-bit ADCs, and have been demonstrated to work successfully only on channels with a small number of multi-path clusters in their geometric channel model representation \cite{mo2014channel,mo2017channel,myers2020low}.

In this paper, we describe an unsupervised learning technique to perform channel estimation from a small number of one-bit quantized pilot measurements using deep generative priors. This technique was introduced in \cite{balevi2020high,doshi2020compressed} to perform full-resolution channel estimation from a limited number of pilots. However, the prior work in \cite{balevi2020high} assumed an antenna spacing of $\lambda_c/10$ (where $\lambda_c = c/f_c$ and $f_c$ is the carrier frequency) in place of the conventional $\lambda_c/2$ to generate high spatial correlation in channel realizations and also trained a deep generative model and evaluated its performance only on channels with a strong LOS component. While we addressed the aforementioned limitations in the context of full-resolution channel estimation in \cite{doshi2022over}, in this paper, we will additionally extend the framework to heavily quantized channel estimation. 

We first describe training a deep generative model using a Generative Adversarial Network (GAN) to output beamspace channel realizations, followed by the training of a single conditional generative model to output a range of LOS and NLOS channel realizations. Subsequently, we will perform channel estimation from compressive one bit pilot measurements by optimizing the input vector to a pre-trained deep generative model with the objective of maximizing the correlation between the generator output and pilot measurements.

\section{System Model}
Consider a single user setup with a transmitter and receiver having $N_t$ and $N_r$ antennas respectively. 
We want to estimate the downlink (DL) narrowband mmWave MIMO channel matrix $\mathbf{H} \in \mathbb{C}^{N_r \times N_t}$ from one-bit quantized received pilot signals $\mathbf{Y}$. Denote the hybrid precoder by $\mathbf{F} \in \mathbb{C}^{N_t \times N_s}$, and the hybrid combiner by $\mathbf{W} \in \mathbb{C}^{N_r \times N_s}$, with $N_s$ being the number of data streams that can be transmitted. 
Consequently, the transmitted pilot symbols $\mathbf{S} \in \mathbb{C}^{N_s \times N_p}$ are received as
\begin{equation} \label{eq:sys_model}
    \mathbf{Y} = \mathcal{Q}_1(\mathbf{W}^H \mathbf{HFS} + \mathbf{W}^H\mathbf{N}), 
\end{equation}
where each element of $\mathbf{N} \in \mathbf{C}^{N_r \times N_p}$ is an independent and identically distributed (i.i.d.) complex Gaussian random variable with mean 0 and variance $\sigma^2$. The operator $\mathcal{Q}_1$ represents one-bit quantization and mathematically is given by 
\begin{equation}
    \mathcal{Q}_1(\cdot) = \mathrm{sign}(\mathrm{Re}(\cdot)) + \mathrm{j}~\mathrm{sign}(\mathrm{Im}(\cdot)).
\end{equation}
Note that while $\mathbf{Y} \in \{\pm 1\}^{N_s \times N_p}$, we want to recover a ``full-resolution" estimate of $\mathbf{H}$. Implicitly assumed in \eqref{eq:sys_model} is a block fading model over $ \geq N_p$ pilot symbols i.e. a new i.i.d. channel realization $\mathbf{H}$ is chosen atmost every $N_p$ time slots. We assume a fully connected phase shifting network \cite{mendez2016hybrid}, and constrain the angles realized by the phase shifters to quantized sets \cite{venugopal2017channel} given by 
\begin{equation}
    \mathcal{A} = \bigg\{0, \frac{2\pi}{2^{N_Q}}, \ldots, \frac{(2^{N_Q}-1)2\pi}{2^{N_Q}} \bigg\},
\end{equation}
where $N_Q$ is the number of quantization bits. We assume $N_{bit,t}$ and $N_{bit,r}$ phase shift quantization bits at the transmitter and receiver respectively, with the quantization sets denoted by $\mathcal{A}_t$ and $\mathcal{A}_r$ respectively. This implies $[\mathbf{F}]_{i,j} = \frac{1}{\sqrt{N_t}} e^{j\psi_{i,j}}$ and $[\mathbf{W}]_{i,j} = \frac{1}{\sqrt{N_r}} e^{j\phi_{i,j}}$ where $\psi_{i,j} \in \mathcal{A}_t$ and $\phi_{i,j} \in \mathcal{A}_r$. Vectorizing \eqref{eq:sys_model} and utilizing the Kronecker product identity $\underline{\mathbf{ABC}} = (\mathbf{C}^{T} \otimes \mathbf{A})\underline{\mathbf{B}}$, we obtain
\begin{equation} \label{eq:kron_sys_model}
    \underline{\mathbf{y}} = \mathcal{Q}_1((\mathbf{S}^T\mathbf{F}^T \otimes \mathbf{W}^H) \underline{\mathbf{H}} + (\mathbf{I}_{N_p} \otimes \mathbf{W}^H) \underline{\mathbf{n}}),
\end{equation}
where $\underline{\mathbf{y}} \in \{\pm 1\}^{N_{\mathrm{s}}N_{\mathrm{p}} \times 1}$, $\underline{\mathbf{H}} \in \mathbb{C}^{N_{\mathrm{r}}N_{\mathrm{t}} \times 1}$ and $\underline{\mathbf{n}} \in \mathbb{C}^{N_{\mathrm{r}}N_{\mathrm{p}} \times 1}$. Since the received signal is 1-bit quantized, channel estimation in the noiseless, full-rank ($N_sN_p = N_tN_r$) setting is also an ill-posed inverse problem. The technique presented in this paper will provide for channel estimation from noisy one-bit pilot measurements with $N_pN_s < N_tN_r$.

\section{Quantized Generative Channel Estimation} \label{sec:qgce}
Deep generative models $\mathbf{G}$ are feed-forward neural networks (NN) that take as input a low dimensional vector $\mathbf{z} \in \mathbb{R}^d$ and output high dimensional matrices $\mathbf{G}(z) \in \mathbb{R}^{c \times l \times w}$, where $c,l$ and $w$ refer to the number of channels, length and width of an image outputted by $\mathbf{G}$ and $d \ll clw$ . Such a model can be trained to take a i.i.d. Gaussian vector $\mathbf{z}$ as input and produce samples from complicated distributions, such as human faces \cite{radford2015unsupervised}. One powerful method for training generative models is using Generative Adversarial Networks (GAN) \cite{goodfellow2014generative}. 

In \cite{balevi2020high}, we developed an algorithm -- Generative Channel Estimation (GCE) -- that utilized compressed sensing using deep generative models \cite{bora2017compressed} to perform MIMO channel estimation. We trained $\mathbf{G}$ to output channel realizations $\mathbf{H}$ from a given distribution, and then utilized $\mathbf{G}$ to recover $\mathbf{H}$ from compressive pilot measurements $\underline{\mathbf{y}}$. However, we experimentally demonstrated that a reduced antenna spacing of $\lambda_c/10$ in the antenna arrays at the transmitter and receiver was key to training $\mathbf{G}$ successfully. We attributed this to the high spatial correlation generated in channel realizations by such an antenna spacing, making it easier for $\mathbf{G}$ to learn the underlying channel distribution. 

In this paper, we will utilize the following key insight to output channel realizations with the conventional and realistic $\lambda_c/2$ antenna spacing: \textit{beamspace representation of mmWave MIMO channels have high spatial correlation due to clustering in the angular domain.} To be precise, assuming uniformly spaced linear arrays at the transmitter and receiver, the array response matrices are given by the unitary DFT matrices $\mathbf{A}_{\mathrm{T}} \in \mathbb{C}^{N_{\mathrm{t}} \times N_{\mathrm{t}}}$ and $\mathbf{A}_{\mathrm{R}} \in \mathbb{C}^{N_{\mathrm{r}} \times N_{\mathrm{r}}}$ respectively. Then, we can represent $\mathbf{H}$ as
\begin{equation} \label{eq:virtual_channel_rep}
\begin{aligned}
    \mathbf{H} &= \mathbf{A}_{\mathrm{R}}\mathbf{H}_\mathrm{v}\mathbf{A}_{\mathrm{T}}^H \\
    \underline{\mathbf{H}} &= ((\mathbf{A}_{\mathrm{T}}^H)^T \otimes \mathbf{A}_{\mathrm{R}})\underline{\mathbf{H}_\mathrm{v}}.
\end{aligned}
\end{equation}
We will train $\mathbf{G}$ to output samples of $\mathbf{H}_{\mathrm{v}}$ i.e. $\mathbb{P}_{\mathbf{G}}$ converges to $\mathbb{P}_{\mathbf{H}_{\mathrm{v}}}$ as GAN training progresses. 

Subsequently, in order to adapt GCE to quantized channel estimation, we propose a new empirical optimization objective, drawing inspiration from the loss function proposed in \cite{Qiu19WeiQiu} for robust one-bit recovery using deep generative networks. Given a trained generator $\mathbf{G}$ and pilot measurements $\mathbf{\underline{y}}$ as defined in \eqref{eq:kron_sys_model}, we will solve the following optimization problem
\begin{equation} \label{eq:gan_eq_rep_obq}
    \mathbf{z}^* = \underset{\mathbf{z} \in \mathbb{R}^d} {\mathrm{arg\ max\ }} \sum_{i=1}^{N_{\mathrm{p}}N_{\mathrm{s}}} \underline{\mathbf{y}}[i]\langle\mathbf{A}_\mathrm{sp}[i],\underline{\mathbf{G}}(z)\rangle
\end{equation}
where $\mathbf{A}_{\mathrm{sp}} = (\mathbf{A}_{\mathrm{T}}^H\mathbf{F}\mathbf{S})^T \otimes \mathbf{W}^H\mathbf{A}_{\mathrm{R}}$. This heuristically designed loss function attempts to maximize the correlation between $\underline{\mathbf{y}}$ (which is constrained to a vector with entries $\pm 1$ for one-bit quantization) and $\mathbf{A}_\mathrm{sp} \underline{\mathbf{G}}(z)$. The summation in \eqref{eq:gan_eq_rep_obq} should should be interpreted as the sum over the real and imaginary parts, separately,
\begin{eqnarray}
 &&\sum_{i=1}^{N_{\mathrm{p}}N_{\mathrm{s}}} \mathrm{Re}(\underline{\mathbf{y}}[i])\mathrm{Re}(\langle\mathbf{A}_\mathrm{sp}[i],\underline{\mathbf{G}}(z)\rangle) + \\ \nonumber &&\sum_{i=1}^{N_{\mathrm{p}}N_{\mathrm{s}}} \mathrm{Im}(\underline{\mathbf{y}}[i])\mathrm{Im}(\langle\mathbf{A}_\mathrm{sp}[i],\underline{\mathbf{G}}(z)\rangle)
\end{eqnarray}
The beamspace channel estimate is then given by $\mathbf{H}_\mathrm{v,est} = \mathbf{G}(\mathbf{z}^*)$. The performance metric used to assess the quality of $\mathbf{H}_\mathrm{v,est}$ is the normalized mean square error (NMSE), defined as 
\begin{equation} \label{eq:NMSE}
    \text{NMSE} = \mathbb{E}\left[\frac{||\mathbf{H}_\mathrm{v}- \kappa\mathbf{H}_\mathrm{v,est}||_2^2}{||\mathbf{H}_\mathrm{v}||_2^2}\right],
\end{equation}
where $\kappa = \mathrm{argmin}||\mathbf{H}_\mathrm{v}- \kappa\mathbf{H}_\mathrm{v,est}||_2^2$ for a given $\mathbf{H}_\mathrm{v}$ and $\mathbf{H}_\mathrm{v,est}$. 

\section{GAN architectures}
In this section, we will present two different GAN architectures - (i) Wasserstein GAN with Gradient Penalty (WGAN-GP) and (ii) Conditional Wasserstein GAN (CWGAN). We will present both architectures in the context of narrowband MIMO channel generation. Moreover, we will assume that the generator in all GAN architectures will output the beamspace MIMO channel representation given by \eqref{eq:virtual_channel_rep}, in accordance with the QGCE framework outlined in Section \ref{sec:qgce}.

\subsection{Wasserstein GAN with Gradient Penalty} \label{subsec:wgan_gp}

A Wasserstein Generative Adversarial Network (WGAN) \cite{arjovsky2017wasserstein} consists of two deep neural networks - a generator $\mathbf{G}(.;\theta_g)$ and a critic $\mathbf{D}(.;\theta_d)$ - whose weights are optimized so as to solve the following min-max problem:
\begin{equation} \label{eq:wgan}
    \underset{\mathbf{G}}{\mathrm{min}}~\underset{\mathbf{D} \in \mathcal{D}}{\mathrm{max}} ~ \mathbb{E}_{\mathbf{x}\sim \mathbb{P}_{r}(\mathbf{x})} \mathbf{D}(\mathbf{x}) 
    - \mathbb{E}_{z\sim \mathbb{P}_{\mathbf{z}}(\mathbf{z})}\mathbf{D}(\mathbf{G}(\mathbf{z})),
\end{equation}
where $\mathcal{D}$ is the set of 1-Lipschitz functions and $\mathbb{P}_{r}$ is the data distribution. We also denote the output distribution of the generator by $\mathbb{P}_{\mathbf{G}} = \mathbb{P}_{\mathbf{z}}(\mathbf{z}) (\nabla_{\mathbf{z}} \mathbf{G}(\mathbf{z}))^{-1}$. The original GAN \cite{goodfellow2014generative} is famously known to suffer from mode collapse \cite{srivastava2017veegan}, i.e. $\mathbb{P}_{\mathbf{G}}$ collapses to a delta function centered around the mode of the input data distribution. In \cite{arjovsky2017wasserstein}, they attribute this behaviour to the use of the KL (Kullback-Leibler) or JS (Jensen-Shannon) divergence during training, and instead propose using the Wasserstein-1 distance to improve robustness to mode collapse. 

WGAN with Gradient Penalty (WGAN-GP) \cite{gulrajani2017improved} improves the performance of WGAN by incorporating a penalty on the gradient norm for random samples $\hat{\mathbf{x}} \sim \mathbb{P}_{\hat{\mathbf{x}}}$ as a soft version of the Lipschitz constraint. In case of WGAN, the Lipschitz constraint $\mathbf{D} \in \mathcal{D}$ in \eqref{eq:wgan} is enforced by clipping $\theta_d$ to be between $(-\tau,\tau)$, where $\tau$ is the clipping constant. In case of WGAN-GP, the Lipschitz constraint is enforced by adding
\begin{equation}
    L_{\mathrm{GP}}(\theta_d) = \mathbb{E}_{\hat{\mathbf{x}}\sim\mathbb{P}_{\hat{\mathbf{x}}}} \big[ (||\nabla_{\hat{\mathbf{x}}} \mathbf{D}(\hat{\mathbf{x}};\theta_d)||_2 - 1)^2 \big]
\end{equation}
to the objective in \eqref{eq:wgan}, where $\hat{\mathbf{x}}$ are points sampled uniformly along straight lines joining pair of points sampled from the data distribution $\mathbb{P}_r$ and the generator distribution $\mathbb{P}_{\mathbf{G}}$, since enforcing the gradient penalty over all possible inputs to $\mathbf{D}$ is intractable \cite{gulrajani2017improved}. A unified algorithm capturing the training of both WGAN and WGAN-GP in the context of channel generation is outlined in Algorithm~\ref{alg:WGAN_GP_training} by utilizing an indicator $\mathbbm{1}_{\mathrm{GP}}$ to indicate if WGAN-GP was chosen or not. 
\begin{algorithm} 
\DontPrintSemicolon
\SetAlgoHangIndent{0pt}
\setstretch{1}
\For{number of training iterations}{
 \For{$n_{\mathrm{d}}$ iterations}{
    Sample minibatch of $m$ beamspace channel realizations $\{\mathbf{H}_{\mathrm{v}}^{(i)}\}_{i=1}^{m} \sim \mathbb{P}_{\mathbf{H}_{\mathrm{v}}}$, latent variables $\{\mathbf{z}^{(i)}\}_{i=1}^{m} \sim \mathbb{P}_z$ and random numbers $\{\epsilon^{(i)}\}_{i=1}^{m} \sim U[0,1]$.\;
    $\tilde{\mathbf{H}}_\mathrm{v} = \mathbf{G}(\mathbf{z};\theta_g)$.\;
    $\hat{\mathbf{H}}_\mathrm{v} = \epsilon  \mathbf{H}_{\mathrm{v}} + (1-\epsilon)  \tilde{\mathbf{H}}_\mathrm{v}$.\;
    $L(\theta_d) = \frac{1}{m} \sum_{i=1}^{m} \mathbf{D}(\tilde{\mathbf{H}}_\mathrm{v}^{(i)};\theta_d) - \mathbf{D}(\mathbf{H}_{\mathrm{v}}^{(i)};\theta_d) + \beta \mathbbm{1}_{\mathrm{GP}} ( || \nabla_{\hat{\mathbf{H}}_\mathrm{v}^{(i)}} D(\hat{\mathbf{H}}_\mathrm{v}^{(i)};\theta_d) ||_2 - 1)^2$\;
    $\theta_d = \theta_d - \gamma \mathrm{RMSProp}(\nabla_{\theta_d} L(\theta_d))$\;
    $\theta_d = \mathbbm{1}_{\mathrm{GP}} \theta_d + (1 - \mathbbm{1}_{\mathrm{GP}}) \mathrm{clip}(\theta_d,-\tau,\tau)$
    }
    Sample minibatch of $m$ latent variables $\{\mathbf{z}^{(i)}\}_{i=1}^{m} \sim \mathbb{P}_z$.\;
    $L(\theta_g) = \frac{1}{m} \sum_{i=1}^{m} -\mathbf{D}(\mathbf{G}(\mathbf{z}^{(i)};\theta_g))$\;
    $\theta_g = \theta_g - \gamma \mathrm{RMSProp}(\nabla_{\theta_g} L(\theta_g))$\;
    }
\caption[caption]{Wasserstein GAN} \label{alg:WGAN_GP_training}
\end{algorithm}

\subsection{Conditional Wasserstein GAN} \label{subsec:cwgan_gp}
The WGAN in \cite{balevi2020high} was trained on channel realizations drawn from a single distribution which was characterized by a very strong Line-of-Sight (LOS) component. Moreover, training on such a channel distribution provides no indication of the generator's ability to learn more complex multi-path channels. In this section, we will present a Conditional WGAN (CWGAN) design that will have the ability to be trained on channel realizations drawn from a plurality of distributions, each yielding channel realizations with varying degrees of approximate sparsity in the beamspace domain.

We combine the architecture of Conditional GAN \cite{mirza2014conditional} with the training procedure of WGAN outlined in Algorithm~\ref{alg:WGAN_GP_training} to develop CWGAN. We assume that we are provided with a binary label $\chi$ indicating whether the channel we are trying to estimate is LOS ($\chi=1$) or NLOS ($\chi=0$). The condition $\chi$ is then passed through a learnable Embedding layer, that embeds an integer as a high dimensional vector, followed by a linear and reshaping layer that has output dimensions $(c_{\chi},l,w)$. This embedded output is then appended along the channel dimension to $\mathrm{Linear}(\mathbf{z})$ of shape $(c_z,l,w)$ to yield an input of size $(c_z+c_{\chi},l,w)$ that is passed through the remaining deep convolutional generative network. A similar procedure is followed while inputting $\chi$ to the critic.

We now need to modify the WGAN training procedure outlined in Algorithm~\ref{alg:WGAN_GP_training} to incorporate the conditional input. To this end, we simply sample $\{\mathbf{H}_{\mathrm{v}}^{(i)},\chi^{(i)}\}_{i=1}^{m} \sim \mathbb{P}_{\mathbf{H}_{\mathrm{v}}}$ and utilize $\{\chi^{(i)}\}_{i=1}^{m}$ as input to $\mathbf{G}$ for computing $\tilde{\mathbf{H}}_{\mathrm{v}}$ and as input to $\mathbf{D}$ for computing $L(\theta_d)$. Despite $\{\chi^{(i)}\}_{i=1}^{m}$ being input to $\mathbf{D}$, the derivative $\nabla_{\hat{\mathbf{H}}^{(i)}_\mathrm{v}}$ continues to remain only w.r.t $\hat{\mathbf{H}}^{(i)}_\mathrm{v}$. Subsequently, we randomly sample $m$ labels in $\{0,1\}$ from a Bernoulli$(0.5)$ distribution and utilize these as the conditional input to both $\mathbf{G}$ and $\mathbf{D}$ for computing $L(\theta_g)$.

\section{Results \& Discussion} \label{sec:results}

\subsection{Data Generation \& Preprocessing}
Channel realizations have been generated using the 5G Toolbox in MATLAB in accordance with the 3GPP specifications TR 38.901 \cite{3gpp.38.901}, consisting of an equal number of realizations of all categories of CDL channels i.e. CDL-A,B,C (which are NLOS) and CDL-D,E (which are LOS). The channel simulation parameters are summarized in Table~\ref{tab:channel_param}. We assume a narrowband block fading model in this paper.
\begin{table}
	\caption[Simulation Parameters]{Simulation Parameters}
	\label{tab:channel_param}
	\begin{tabular}{ |p{3cm}|p{2.5cm}|}
		\hline
		$N_{\mathrm{t}}$ & 64 \\
		\hline
		$N_{\mathrm{r}}$ & 16 \\
		\hline
        Antenna Array Type & ULA\\
 		\hline
 		Antenna Spacing & $\lambda/2$\\
		\hline
		Carrier Frequency & 40 GHz\\
		\hline
        Dataset Size & Train - $6000 \times 5$\\& Test - $50 \times 5$ \\\hline
	\end{tabular}\centering
\end{table}

The generator output $\mathbf{G}(z)$ and discriminator input are of size $(2,N_t,N_r)$, where the first dimension allows us to stack the real and imaginary parts. Based on empirical evidence that a GAN is unable to learn mean-shifted distributions \cite{srivastava2017veegan}, it is important to normalize the data used to train a GAN. Given a beamspace channel realization $\mathbf{H}_{\mathrm{v}}$, $\mu[i,j] = \mathbb{E}\big[\mathbf{H}_{\mathrm{v}}[i,j]\big]$, $\mathrm{Re}(\sigma[i,j]) = \big(\mathbb{E}\big[(\mathrm{Re}(\mathbf{H}_{\mathrm{v}}[i,j] - \mu[i,j]))^2\big]\big)^{0.5}$ and $\mathrm{Im}(\sigma[i,j]) = \big(\mathbb{E}\big[(\mathrm{Im}(\mathbf{H}_{\mathrm{v}}[i,j] - \mu[i,j]))^2\big]\big)^{0.5}$, we normalize the matrix element-wise as
\begin{align} \label{eq:normalize}
    \mathrm{Re}(\mathbf{H}_{\mathrm{v}}[i,j]) \leftarrow \frac{\mathrm{Re}(\mathbf{H}_{\mathrm{v}}[i,j] - \mu[i,j])}{\mathrm{Re}(\sigma[i,j])} \\  \mathrm{Im}(\mathbf{H}_{\mathrm{v}}[i,j]) \leftarrow \frac{\mathrm{Im}(\mathbf{H}_{\mathrm{v}}[i,j] - \mu[i,j])}{\mathrm{Im}(\sigma[i,j])}.
\end{align}
In lieu of \eqref{eq:normalize}, the operations $\mathbf{G}(\cdot)$ and $\mathbf{D}(\cdot,\chi)$ will implicitly be used to denote $\mathrm{SN}^{-1}(\mathbf{G}(\cdot))$ and $\mathbf{D}(\mathrm{SN}(\cdot),\chi)$ respectively throughout the paper without exception. Here $\mathrm{SN}(\cdot)$ denotes the operation of \textit{stacking} the real and imaginary part followed by \textit{normalization} using $\{\mu[i,j],\sigma[i,j]\}_{i,j}$ and then $\mathrm{SN}^{-1}(\cdot)$ corresponds to \textit{unnormalization} followed by \textit{unstacking} to generate a complex-valued output.

\subsection{Neural Network Architectures \& Training Hyperparameters} \label{subsec:nn_arch}
The generator and discriminator employed in the Wasserstein GAN are Deep Convolutional NNs. While the discriminator architecture was adopted from \cite{arjovsky2017wasserstein}, the generator was fine-tuned to improve its ability to learn the underlying probability distribution. The generator $\mathbf{G}$ takes an input $z \in \mathbb{R}^d$, passes it through a dense layer with output size $128N_{\mathrm{t}}N_{\mathrm{r}}/16$, and reshapes it to an output size of $(N_{\mathrm{t}}/4,N_{\mathrm{r}}/4,128)$. This latent representation is then passed through $k=2$ layers, each consisting of the following units: $2 \times 2$ upsampling, 2D Convolution with a kernel size of 4 and Batch Normalization. All BatchNorm2D layers have $\mathrm{momentum}=0.8$ \cite{arjovsky2017wasserstein} and Conv2D layers have $\mathrm{bias} = \mathrm{False}$\footnote{Code available at https://github.com/akashsdoshi96/obq-gan-mimo-ce.}. We utilize $d = 65$ for $\mathbf{z} \in \mathbb{R}^d$ (refer Appendix B in \cite{doshi2022over} for an empirical justification). 

In order to extend the generator and critic architectures to the conditional setting, we employ an Embedding$(2,10)$ layer in both. This layer learns a $10$-dimensional embedding for $\chi=0$ and $\chi=1$. Subsequently, $\mathbf{G}$ passes this embedding through Linear$(10,N_tN_r/16)$ and Reshape$(1,N_t/4,N_r/4)$ before concatenating it to Linear$(\mathbf{z})$ of size $(127,N_t/4,N_r/4)$. 

In Algorithm~\ref{alg:WGAN_GP_training}, we set $n_{\mathrm{d}}=5,~\beta=10,~\tau=0.01$ and $\gamma=0.00005$ \cite{gulrajani2017improved}\cite{arjovsky2017wasserstein}. We utilize a minibatch size of $m = 200$ in all GAN training. For performing QGCE, we utilize an Adam \cite{kingma2014adam} optimizer with a step size $\eta = 0.1$ and iteration count $500$. We also determined empirically that resetting the RMSProp optimizer for the critic at every training iteration improved the performance of Algorithm \ref{alg:WGAN_GP_training}. We will utilize $N_s = 16$, $N_p = 25$, $N_{bit,t} = 6$ and $N_{bit,r} = 2$. Note that $N_sN_p < N_tN_r$, hence the sensing matrix $\mathbf{A}_{\mathrm{sp}}$ is not full rank, and channel estimation is an ill-posed inverse problem even in the absence of quantization.

\subsection{Baselines}
\subsubsection{BG-GAMP} We utilize the Generalized Approximate Message Passing (GAMP) algorithm proposed in \cite{mo2014channel} \cite{mo2017channel} as a compressed sensing baseline for 1-bit quantized channel estimation. Specifically, \cite{mo2017channel} models the angular domain coefficients of the signal to be recovered - in this case, the beamspace channel - as a Bernoulli-Gaussian (BG) mixture random variable and uses AMP to compute approximately the MMSE estimates of the channel coefficients. We also tune the sparsity hyperparameter in accordance with the approximate beamspace sparsity of each channel model. To be precise, \cite{mo2017channel} defines the channel sparsity rate $1 - \lambda_0$ as the ratio of the number of non-zero elements in $\mathbf{H}_{\mathrm{v}}$ and $N_tN_r$. Based on the beamspace CDL channel representations, we use the following estimates for $(1-\lambda_0)N_tN_r$ while implementing BG-GAMP : CDL-A (20), CDL-B,C (50), CDL-D,E (5).

\subsubsection{GCE} 
Given noisy un-quantized pilot measurements $\mathbf{\underline{y}}$ and $\lambda_{\mathrm{reg}} = 0.001$, GCE \cite{doshi2022over} recovers the channel estimate $\mathbf{H}_\mathrm{v,est} = \mathbf{G}(\mathbf{z}^*)$, where $\mathbf{z}^*$ is given by
\begin{equation} \label{eq:gce}
    \mathbf{z}^* = \underset{\mathbf{z} \in \mathbb{R}^d} {\mathrm{arg\ min\ }} \hspace{0.05 in} ||\underline{\mathbf{y}} - \mathbf{A}_{\mathrm{sp}} \mathbf{\underline{G}}(\mathbf{z})||_{2}^2 + \lambda_{\mathrm{reg}} ||\mathbf{z}||_{2}^2,
\end{equation}

\subsection{Results}
\begin{figure*}
    \centering
    \includegraphics[width=5.7in]{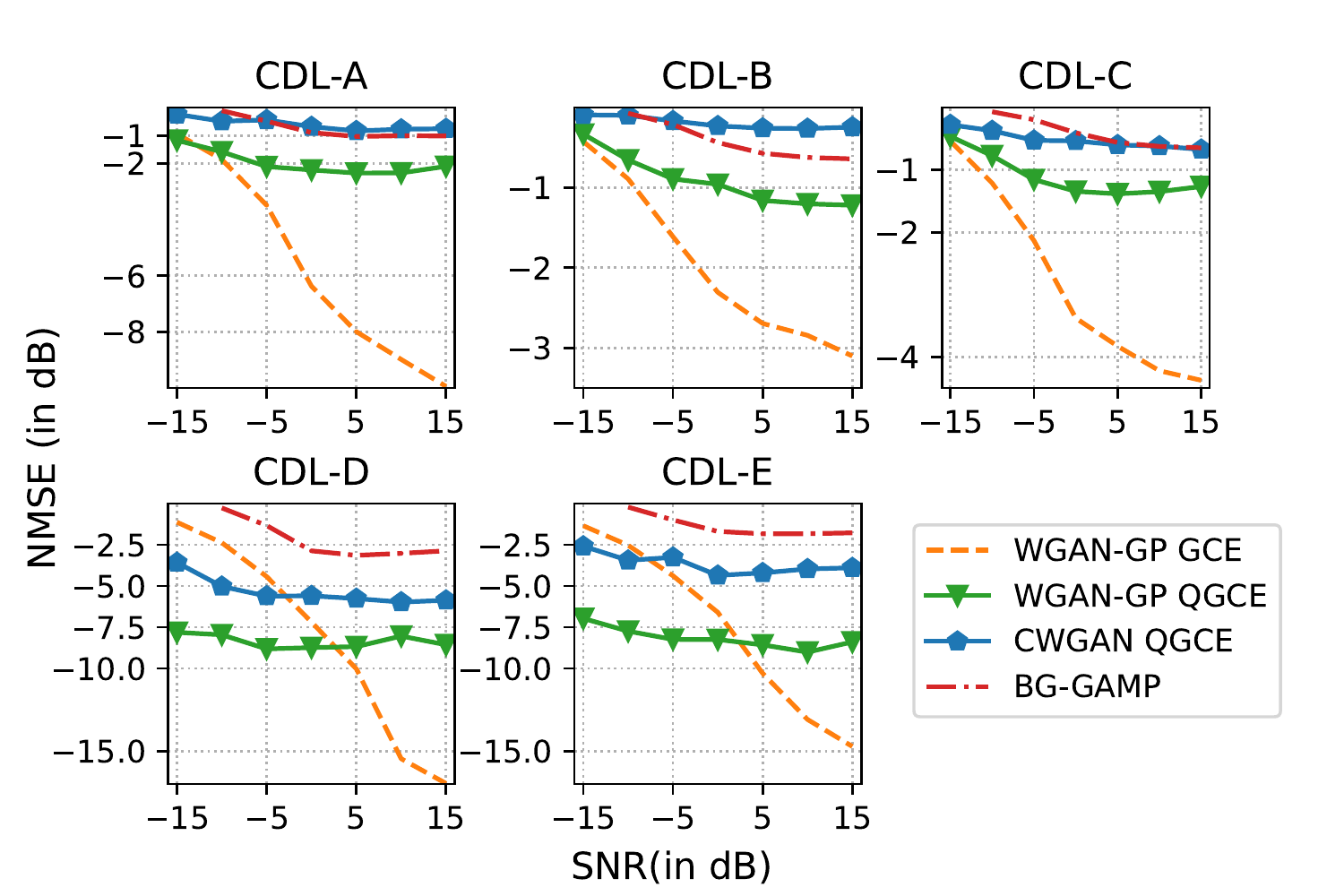}
    \caption{NMSE vs SNR for WGAN-GP GCE \& QGCE and CWGAN QGCE. For reference, BG-GAMP baselines have also been plotted.}
    \label{fig:nmse_snr}
\end{figure*}

In accordance with the WGAN-GP model developed in Section \ref{subsec:wgan_gp}, we design a separate generator for each of the five CDL channel models by training a WGAN-GP ($\mathbbm{1}_{\mathrm{GP}} = 1$) using Algorithm~\ref{alg:WGAN_GP_training} for 60,000 training iterations. We also design a single conditional generative model by training a CWGAN ($\mathbbm{1}_{\mathrm{GP}} = 0$) as outlined in Section~\ref{subsec:cwgan_gp} for 100,000 training iterations. In both cases, we extract the final trained generator $\mathbf{G}$ and perform QGCE at varying SNR to plot NMSE vs SNR, as shown in Fig. \ref{fig:nmse_snr}. 

Clearly, the individually trained WGAN-GP outperforms BG-GAMP, by 0.5 dB in CDL-B and C, 1 dB in CDL-A and 5 dB for CDL-E and D. We also observe that across CDL channel models, the performance of QGCE is consistently $\mathrm{B} < \mathrm{C} < \mathrm{A} < \mathrm{E} < \mathrm{D}$.
This is in agreement with the decreasing number of rays/clusters and the increasing magnitude of the LOS component in $\mathbf{H}_{\mathrm{v}}$ as we go from left to right (refer Table 7.7.1 of \cite{3gpp.38.901} for the precise channel profiles). It is important to note that the NMSEs obtained using QGCE, going as low as -7.7 dB for CDL-D and CDL-E , have been obtained using only a fraction -- $\alpha = N_pN_s/N_tN_r = 0.4$ -- of the pilot symbols that would have been required for full-rank channel estimation in the \textit{absence} of quantization. For comparison, one-bit quantized channel estimation algorithms in \cite{mo2017channel} and \cite{myers2020low} use $\alpha = 8$ to perform GAMP-based CE. Aside from the excessive training overhead, such methods also implicitly assume that the channel does not change over a large number of pilot symbols, rendering them inapplicable in the presence of UE mobility.

As expected, we see a degradation in NMSE compared to full-resolution GCE, ranging from $\sim 8$ dB for CDL-A and D to $2$ dB for CDL-B at an SNR of 15 dB. However, it is interesting to observe that QGCE outperforms GCE at SNR $\leq 0$ dB for CDL-D and E. This suggests that the QGCE optimization in \eqref{eq:gan_eq_rep_obq} is more robust to noise than the GCE optimization in \eqref{eq:gce}, and a weighted objective combining \eqref{eq:gan_eq_rep_obq} and \eqref{eq:gce} could be used to improve the performance of full-resolution channel estimation at low SNRs.

On switching to a single CWGAN model, we observe that the gain in NMSE over BG-GAMP is reduced. The LOS channel models CDL-D and E still outperform BG-GAMP, but the NLOS channel models only acheive performance competitive with BG-GAMP. 
The performance degradation in CWGAN as compared to the individually trained WGAN-GP generative models can be attributed to the usage of a simple LOS/NLOS label to distinguish between the different channel modalities as well as the inability of the convolutional architecture of the generator $\mathbf{G}$ to learn the NLOS channel models that have ``richer" beamspace representations. 

At the same time, it should be noted that the BG-GAMP baseline was adapted to each CDL channel model by a careful tuning of the sparsity rate for CDL A-E. Such sparsity rates cannot be obtained in practice from pilot measurements alone. We initially considered the usage of EM-BG-GAMP as described in \cite{mo2017channel}, where the Expectation Maximization -- EM -- step would be responsible for the automated tuning of the sparsity and noise variance estimates, however we were unable to obtain any reasonable NMSE for one-bit quantized pilot measurements, even with a higher value of $N_p$. A possible reason for this could be that \cite{mo2017channel} only tested channels with a small number of multi-path clusters ($\leq 4$) in their geometric channel model representation, while the CDL channel models contain upto 23 clusters.

\begin{figure*}
    \centering
    \includegraphics[width=5.5in]{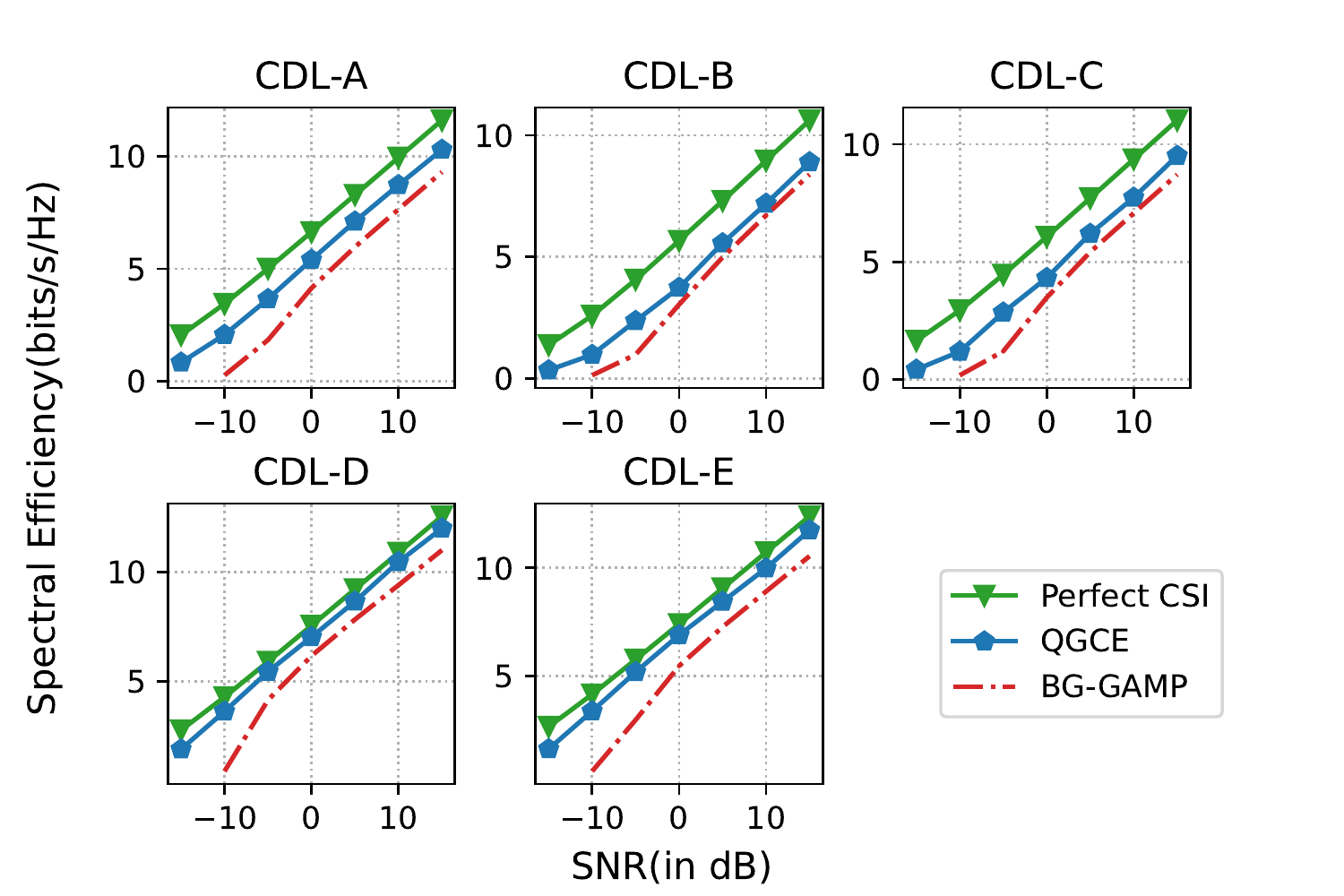}
    \caption{Spectral Efficiency vs SNR for WGAN-GP QGCE. For reference, BG-GAMP and Perfect CSI baselines have also been plotted.}
    \label{fig:nmse_capacity}
\end{figure*}

To compute the NMSE in \eqref{eq:NMSE}, note that we utilize a seemingly genie-aided scaling factor $\kappa$, since both the one-bit quantized pilot measurements $\mathbf{y}$ as well as the correlation-based optimization objective in \eqref{eq:gan_eq_rep_obq} do not provide for optimal scaling of the reconstructed channel. In order to verify that the channel estimate $\mathbf{H}_{\mathrm{est}} = \mathbf{A}_{\mathrm{R}}\mathbf{G}(\mathbf{z}^*)\mathbf{A}_{\mathrm{T}}^H$ from WGAN-GP based QGCE is in fact ``better" than the estimate obtained from BG-GAMP, we perform a simple achievable rate computation. Utilizing the SVD of $\mathbf{H}_{\mathrm{est}} = \mathbf{U}\mathbf{\Sigma}\mathbf{V}^H$, we set the optimal precoding and combining vector  as the first column of $\mathbf{V}$ and $\mathbf{U}$ respectively. The spectral efficiency is then given by 
\begin{equation}
    C(\mathrm{SNR};\mathbf{H}) = \log_2(1 + \mathrm{SNR}||\mathbf{U}[1]^H \mathbf{H} \mathbf{V}[1]||^2)
\end{equation}
The spectral efficiency is plotted as a function of SNR in Fig. \ref{fig:nmse_capacity}. One can observe the clear correspondence with the NMSE in Fig. \ref{fig:nmse_snr}. For e.g., $C(\mathrm{SNR})$ for CDL-D and E almost matches the Perfect CSI curve, unlike CDL-A,B and C, since CDL-D and E acheive NMSEs as low as -8 dB. Similarly, $C(\mathrm{SNR})$ for CDL-B and C has the least improvement over BG-GAMP, which is again consistent with the small $\sim 0.5$ dB improvement in NMSE over BG-GAMP in Fig. \ref{fig:nmse_snr}. 

\section{Conclusions and Future Directions}
Channel estimation in mmWave MIMO using one-bit quantized pilot measurements typically requires a large number of pilot measurements ($N_p \gg N_tN_r/N_s$) in order to recover a channel estimate with low NMSE. In this paper, we demonstrate how a deep generative prior $\mathbf{G}$, trained using Wasserstein GAN, can be used to perform channel estimation from a limited number of pilot measurements ($N_p < N_tN_r/N_s$) by optimizing the input vector $\mathbf{z}$ to a deep generative model $\mathbf{G}$ with the objective of maximizing the correlation between the quantized pilot measurements $\mathbf{y}$ and the estimated transmit signal $\mathbf{A}_{\mathrm{sp}}\mathbf{G}(\mathbf{z})$. Our results indicate that a carefully tuned generative prior significantly outperforms state-of-the-art baselines such as BG-GAMP, while a single conditional generative model outperforms BG-GAMP on LOS channel models and achieves competitive results on NLOS channel models.

A key shortcoming of our approach is the need for clean channel realizations to train the WGAN. While techniques such as Ambient GAN \cite{bora2018ambientgan} can be used to train WGAN from noisy un-quantized pilot measurements, the usage of one-bit ADCs destroys the invertibility of the function mapping the probability density $p_{\mathbf{H}_{\mathrm{v}}}(\mathbf{H}_{\mathrm{v}})$ to $p_{\mathbf{y}}(\mathbf{y})$, rendering Ambient GAN inapplicable. Hence \textit{training} a GAN using noisy quantized pilot measurements should be investigated. Additionally, instead of a simple LOS/NLOS label, we could train $N$ GANs -- for example $N=3$ for low, medium and high levels of beamspace sparsity -- and then learn a classifier that will indicate which generative model to utilize. 

\bibliographystyle{IEEEtran}
\bibliography{bibtex.bib}

\end{document}